\input phyzzx
\sequentialequations
\overfullrule=0pt
\tolerance=5000
\nopubblock
\twelvepoint

\line{\hfill  IASSNS-HEP 97/48}
\line{\hfill hep-th/9706014 }
\line{\hfill June 1997}

\titlepage
\title{Mass Splittings From Symmetry Obstruction}
\vskip.6cm
\author{Lorenzo Cornalba\foot{cornalba@princeton.edu}}
\vskip.2cm
\centerline{{\it Department of Physics}}
\centerline{{\it Joseph Henry Laboratories}}
\centerline{{\it Princeton University}}
\centerline{{\it Princeton, NJ 08544, USA}}
\vskip.4cm
\author{Frank Wilczek\foot{Research supported in part by DOE grant
DE-FG02-90ER40542.~~~wilczek@sns.ias.edu}}
\vskip.2cm
\centerline{{\it School of Natural Sciences}}
\centerline{{\it Institute for Advanced Study}}
\centerline{{\it Olden Lane}}
\centerline{{\it Princeton, N.J. 08540}}
 
\endpage

\abstract{The long-range fields associated with 
non-abelian vortices generally obstruct full realization, in the
spectrum,
of the
symmetry of the ground state.  
In the context of 2+1 dimensional field theories, we
show how this effect manifests itself concretely in altered conditions
for the angular momentum and in the energy spectrum.  A particularly
interesting case is supersymmetry, which is obstructed by the
gravitational effect of any mass.}

\endpage
 
%references on obstructed symmetry idea
%
\REF\monopoles{A. Balachandran, G. Marmo, M. Mukunda, J. Nilsson,
E. Sudarshan, and F. Zaccaria,  {\it Phys. Rev. Lett.} {\bf 50}, 1553
(1983); S. Coleman and P. Nelson, {\it Nucl. Phys.} {\bf B237}, 1 (1984).}

\REF\vortices{A. Schwarz, {\it Nucl. Phys.} {\bf B208}, 141 (1982); 
M. Alford, K. Benson, S. Coleman, J. March-Russell, and F. Wilczek
{\it Phys. Rev. Lett.} {\bf 64}, 1632 (1990); {\it Nucl. Phys.} {\bf
B349}, 414 (1991).  There has been very
substantial generalization and mathematical development of the ideas;
see for example F. Bais, P. van Driel, and M. de Wild Propitius {\it
Nucl. Phys.} {\bf B393} 547 (1993).}

\REF\supergravity{E. Witten, {\it Int. J. Mod. Phys.} {\bf A10}, 1247 (1995).}

The existence of fundamental
symmetries that are somehow hidden -- spontaneously broken,
or asymptotic -- is an important theme in modern physical theory.
There is another mechanism for hiding symmetry that has been less
analyzed, and does not seem to have acquired a name, 
though it is known to occur in various contexts.  This phenomenon,
which we propose to call obstructed symmetry, is connected with the
existence of long-range gauge fields (or potentials).  
Symmetry transformations among charged fields 
are implemented, in quantum gauge field theory, by
operators involving an integral over a surface at spatial infinity of
the electric field.  In a non-abelian theory the electric field itself
is gauge covariant, not invariant, 
so that to define the surface integral uniquely one
must have a consistent global definition of the gauge frames over the
surface.  Long-range gauge fields, even if they reduce to locally
trivial gauge potentials, can obstruct such a definition.  In that
case, the usual consequences of the symmetry may not hold [\monopoles
, \vortices , \supergravity ]. 

%additional cosmo. const. references

\REF\cosmo{E. Witten, {\it Mod. Phys. Lett.} {\bf A10}, 2153 (1995).}

\REF\bbd{K. Becker, M. Becker, and A. Strominger {\it Phys. Rev.}
{\bf D51}, 6603 (1995).}

All this is rather abstract, and our first goal in this 
note is just to exemplify (conceptually) tangible consequences of
obstructed symmetry in the simplest possible context, that of charges
bound to non-abelian vortices in 2+1 dimensions.  We show explicitly
how the angular momentum quantization
conditions are modified, depending on internal quantum numbers, in
such a way that the symmetry of the energy spectrum is reduced.  
Essentially the same phenomenon occurs universally in 2+1 dimensional
supergravity, and plays an important role in some recent ideas
concerning the vanishing of the cosmological constant
[\supergravity , \cosmo , \bbd ].  It is
important for these ideas that supersymmetry is not broken, but that
its obstruction leads to fermion-boson splitting.  We show explicitly,
in a special case, how this occurs.

\noindent{\it Non-abelian vortices\/}:

A nonabelian vortex is characterized in regular gauge
by a non-zero matrix vector potential
$a_\theta$ at spatial infinity, and the holonomy
$$
h ~=~ {\cal P} e^{{i\int^{2\pi}_0} a_\theta~ d\theta}~.
\eqn\holonomy
$$
Here ${\cal P}$ denotes path ordering, and $a$ is a matrix in the
adjoint representation.  (More precisely, $h$ is defined only up to
conjugacy, since a gauge transformation $\Lambda(r, \theta )
\rightarrow \Lambda_\infty ( \theta)$ takes 
$h \rightarrow \Lambda_\infty (2\pi ) h \Lambda_\infty (0)^{-1}$.)  
We may suppose, after a non-singular gauge
transformation, that 
$a_\theta$ is a constant matrix.  Given a charged field $\eta$, one
has its covariant derivative
$$
D_\mu \eta ~=~ \partial_\mu \eta + i {\rm T}^a a^{\rm a}_\mu \eta~.
\eqn\covderiv  
$$
Here ${\rm T}^{\rm a}$ are the generators for the representation to which 
$\eta$ belongs, and of course the $a^{\rm a}$ are the internal
components of $a$.  The physical implication of the vector potential
emerges clearly if one diagonalizes ${\rm T}^{\rm a} a^{\rm a}_\mu$
(assumed constant) and considers the partial wave
$\eta (r, \theta ) = e^{il\theta} f(r)$.  In the regular gauge $l$ is
quantized to be an integer, but for the $k^{\rm th}$ eigenvalue
$e_{(k)}$ and eigenvector $\eta_{(k)}$ one has the azimuthal
derivative
$$
D_\theta \eta_{(k)} ~=~ i(l + e_{(k)} ) \eta_{(k)} ~.
\eqn\correction
$$
Thus the angular energy term, which has direct physical significance, 
is proportional to ${(l + e_{(k)})^2\over
r^2}$.  It behaves as if there is a fractional contribution $e_{(k)}$ 
to the
angular momentum.  

Alternatively, by a singular gauge transformation
(with gauge function proportional to $\theta$, and thus having a cut)
one could formally remove $a_\theta$, at the cost of introducing
boundary conditions of the form
$$
\eta (2\pi ) ~=~ \rho (h) \eta (0) 
\eqn\singgauge
$$
on the azimuthal dependence of the field $\eta$ (and thus on the wave
functions of its quanta).  Here of course $\rho$ is the appropriate
unitary representation matrix.  Upon diagonalizing $\rho$ one finds 
conditions $\eta_{(k)} (2\pi ) = e^{i\phi_{(k)}} \eta_{(k)} (0)$ on the wave
functions, in evident notation.  In this formulation the azimuthal
covariant derivative is simply the ordinary spatial azimuthal
derivative, but one has the condition 
$$
l_{(k)} ~=~ {\rm integer}~ +~ {\phi_{(k)}\over 2 \pi } 
\eqn\fracangmom
$$ 
on the partial waves $\eta_{(k)} \propto e^{il\theta} f(r)$.  The form
of the angular energy term, of course, is the same as before, with $e_{(k)}
\equiv \phi_{(k)}/2\pi$.  

Generically states with a given value of $e_{(k)}$ (and the same
spatial wave functions) will be degenerate, but states with different
values of $e_{(k)}$ will not be degenerate, because they see a different
effective Hamiltonian -- or, alternatively, because they
obey a different angular
momentum quantization condition.
Now suppose that the symmetry of the 
ground state is a non-abelian group $K$, and
consider $\kappa ~\in~ K$.  If $\rho$ is a faithful representation, and 
$h\kappa \neq \kappa h$, then $\rho (\kappa )$ will not act within the
spaces of fixed $e_{(k)}$, but will connect states with different
values of $e_{(k)}$.   As we have seen, generically this means that
$\rho (\kappa )$ will connect states of different energy.  Thus the
obstructed symmetry $K$ will not be realized as a symmetry of the
spectrum, but will connect states with different energy.  Only 
members of the 
centralizer group ${\cal C} (h)$ of transformations which commute with $h$
will generate spectral symmetries.   

\REF\alice{A. Schwarz, reference 2.}

As a simple example, consider the Alice string [\alice ].  It arises from the
symmetry breaking $SO(3) \rightarrow SO(2) \times Z_2$ (semidirect
product)
induced by the
vacuum expectation value $T_{11} = T_{22} = {-1\over 2} T_{33} \neq 0$ 
of a traceless symmetric tensor.  The residual symmetries are products
of  rotations
in the 1-2 plane and the parity transformation diagonal $(1, -1, -1)$. 
Ordinarily the representations associated with non-vanishing $SO(2)$
charge $Q$ would fall into two-dimensional multiplets of $SO(2)\times
Z_2$ composed of degenerate states with charges 
$Q$ with $-Q$.  However, in the presence of a vortex with 
holonomy diagonal $(1, -1, -1)$ this doublet breaks up into a
symmetric combination with integer effective angular momentum and an
antisymmetric combination with effectively half-odd integer angular
momentum.  In a generic potential, or specifically for example a
harmonic oscillator potential, these states are not degenerate.  

The Alice string model has the annoying complication that the charge
induces a Coulomb field with logarithmically divergent energy, so
that to set up the problem properly one must consider neutral
configurations, such as particle-antiparticle pairs (and then one gets
involved with the peculiarities of Cheshire charge).   These problems
can be avoided in a different symmetry breaking scheme, where the
symmetry of the ground state is a nonabelian discrete group.  Perhaps
the simplest example is the breaking $SU(2) \rightarrow {\rm unit
quaternions}~$ which can be obtained with a symmetric four-index
spinor having as its only non-vanishing expectation values
$S_{1111} = S_{2222} \neq 0$.  The doublet representation of the unit
quaternions, in the presence of a vortex with holonomy $\sigma_3$, for
example, breaks up into two singlets with integer and half-odd
integer effective angular momentum, much as in the Alice case.

\REF\books{For a general review and a collection of 
reprints on Chern-Simons theory,
see  F. Wilczek, {\it Fractional Statistics and Anyon
Superconductivity}, (World Scientific, Singapore 1990). This is
mostly concerned with abelian theories.  For the nonabelian theory, see
especially following reference.}

\REF\cstopology{E. Witten, {\it Comm. Math. Phys.} {\bf 121}, 351 (1989).}

The effects of obstructed symmetry appear directly only in states
carrying both flux and charge.  In the preceding discussion, these
ingredients were rather artlessly brought together by considering
bound states of pure flux and pure charge.  A more intrinsic form of
the phenomenon occurs in Chern-Simons theories, where the interaction
itself 
naturally associates flux with charge [\books ].  The question then arises
whether splittings are induced for the fundamental quanta, which carry
both charge and flux.  We certainly expect they will, for the
following reason.  In evaluating the self-energy of such a quantum,
one must sum over space-time trajectories where its world line is
self-linked or knotted.  The charge from one part of the trajectory gets
entangled with the flux emanating from another part, and the
propagation is altered, as above.  For non-dynamical sources
(Wilson lines) it is known that the amplitude of a trajectory depends
on its topology; indeed, this fact is at the root of the remarkable
application of non-abelian Chern-Simons theories to topology
[\cstopology ].  It
would be very interesting to do a dynamical calculation along these lines.

\noindent{\it Supersymmetry\/}:

\REF\threedgravity{For a review of 2+1 dimensional gravity, with
extensive references, see R. Jackiw, Planar Gravity, in {\it Diverse
Topics in Theoretical and Mathematical Physics}, (World Scientific,
Singapore, 1995).}

In 2+1 dimensional gravity [\threedgravity ]
the primary effect of a point mass is to
create a conical geometry in its exterior, with a deficit angle $\delta$
proportional to the mass; $\delta = 2\pi Gm$.  
The exterior geometry is then locally flat,
but non-trivial globally.  Indeed, it induces a modification in the
angular momentum quantization for particles in the exterior, similar
in some ways to the non-abelian vortex.  A notable difference is that
whereas the non-abelian vortex induced an additive change in the
quantization condition, change induced by 
the gravitational field of a point mass
is multiplicative. Indeed, from requiring that the azimuthal factor
$e^{il\theta}$ be single-valued as 
$\theta \rightarrow \theta + (2\pi - \delta)$ 
we find $l = {\rm integer}/(1  - {\delta\over 2\pi} )$.  
This difference reflects the
different symmetry of the sources: an additive shift in angular
momentum requires an intrinsic orientation violating the discrete
symmetries P and T, while a multiplicative factor is even under these
transformations.   

The conical geometry presents a global obstruction to defining a
spinor supercharge (see below), so that supersymmetry, even if valid for the
ground state, is obstructed for massive states.  One could therefore
reasonably expect that fermion-boson degeneracy is lifted
[\supergravity , \cosmo , \bbd ], but a
concrete calculation might be welcome.  

Perhaps the simplest way to illuminate the central issue is to
consider what is the condition analogous to \fracangmom\  in this context.
One can remove the effect of the conical geometry by a singular gauge
transformation, in favor of a modified boundary condition.  Here the
covariant derivative should respect parallel transport, so that we
should require that the effect of orbiting the apex of the cone is to
rotate spinors (or vectors) through the appropriate angle, {\it
i.e}. to multiply them by a phase proportional to the spin and angle.
An important subtlety, which arises even in flat space, 
is that one must include a factor -1 for half-odd integer (fermion) fields. 
This factor is essentially equivalent 
to the -1
accompanying fermion loops in Feynman graphs, and its inclusion
implements the normal spin-statistics relation.  With this factor 
$e^{2\pi is}$ taken out, the appropriate condition on wave functions
$\psi_s \propto f(r) e^{il \theta}$ for spin $s$ quanta, 
in order that they transform properly, is 
$$
\psi_s (2\pi - \delta) ~=~ e^{i (2\pi - \delta )l - i \delta s } \psi_s (0) 
\eqn\sbdrycond
$$
leading to 
$$
l ~=~ {1\over 1- {\delta\over 2\pi}} {\rm integer} ~+~ 
 s {{\delta\over 2\pi}\over 1 - {\delta\over 2\pi}} ~.
\eqn\lsquant
$$

\REF\sqed{W. Buch\"uller, S. Love, and R. Peccei, {\it Nucl. Phys.}
{\bf B204}, 429 (1982); P. DiVechhia and V. Schuchhardt, {\it
Phys. Lett.} {\bf 155B}, 427 (1985).}

For our purposes, the crucial aspect of \lsquant\ is simply that the
quantization condition on $l$ becomes, for $\delta \neq 0$,  
$s$-dependent.  This certainly suggests a general mechanism for
splitting states formally (that is, here, locally) related by
supersymmetry.  By forbidding $l=0$ for $s=1/2$, it also provides the
primary obstruction to the construction of well-defined supercharge generators.

To see this mechanism at work concretely, let us 
consider an analogue of the
hydrogen atom.  One can construct a supersymmetric version of quantum
electrodynamics, containing very massive protons and sprotons and
lighter electrons and selectrons, and consider the analogue of the
hydrogen atom [\sqed ].  The main novelty introduced by supersymmetry is
that there is a boson-boson (sproton-selectron) system which mixes
with the fermion-fermion system by photino exchange, and similarly a
boson-fermion/fermion-boson complex.  These two complexes are
related by supersymmetry, and form degenerate multiplets.  

\REF\longer{L. Cornalba and F. Wilczek, paper in preparation.}

This superelectrodynamics can be coupled to
supergravity [\longer ], 
and one can consider gravitational corrections to the
`hydrogen' spectrum. 
One can, in this context, consider the corrections to the effective 
non-relativistic hamiltonian due to one photino exchange. The computation 
[\sqed, \longer ] follows closely the classical derivation of 
the Breit hamiltonian for positronium and takes the form
$$
\Delta {\cal H} ~=~ {-i\over 2m } \bigl [ P_i, V(x)\bigr ] \Sigma_i~,
\eqn\pertH
$$
where $m$ is the electron mass, $V$ is the Coulomb  potential ${e^2\over 
2\pi} \ln r $, and the $\Sigma_i$ are matrices acting in the internal spin 
space of the electron-proton system. 
\pertH\  does not yet include any coupling to gravity.  
We expect the  form of this correction to the 
hamiltonian to be unaltered by gravitational effects, since $\Delta{\cal 
H}$ is a local operator on the wavefunction and gravity in $2+1$ 
dimensions has only global topological effects.  This expectation is 
reinforced by its consistency with the boundary conditions
mentioned above. Indeed the exact form of the matrices 
$\Sigma_i$ implies that, in a conical geometry, the operator 
$\Delta{\cal H}$ becomes hermitian (in a non-trivial way) 
if the boundary condition \sbdrycond is 
satisfied by the components of the wavefunction.  
For the spin 
matrices connect components of the wavefunction with internal spin 
differing by $1$, and the phase acquired by the operator $[ P_i, 
V(x)\bigr ]$ (or simply by $x_i$ after taking the commutator), which
would otherwise spoil hermiticity, is exactly 
compensated by the different, spin-dependent, boundary conditions on
these components.

The main effect of the gravitational field, 
however, arises already in the nominally
spin-independent interaction, which dominates in the low-velocity
(non-{\it special\/} relativistic) limit: the modified condition \lsquant\
for the allowed angular momenta modifies the effective
Schrodinger equation, and splits the spectrum corresponding to
different spin values.  Here we are concerned with the motion of the
electron in the geometry and potential provided by the proton, and take
$\delta = 2\pi GM$.  Clearly supersymmetry is obstructed,
and it is not manifest in the spectrum.  

We have glossed over several technicalities that do not substantially
affect this leading-order calculation, though 
in extending it to higher orders in $e^2$ and 
$m/M$ they would need to be
carefully addressed.  The
most interesting, perhaps, comes from the double-conical geometry
arising when one includes the gravitational fields of both particles.
One then obtains a modified quantization condition on the relative
angular momentum, which takes the form
$$
l_{\rm rel.} ~=~ {{\rm integer}\over 1 - {\delta_{\rm max}\over 2\pi}}
~+~ 
{1\over 2\pi }{(\delta_1 s_2 + \delta_2 s_1)   
\over 1 - {\delta_{\rm max}\over 2\pi}}~,
\eqn\doublecone
$$
where $\delta_i, s_i$ are the deficit angle and spin associated
with particle $i$, and $\delta_{\rm max}$ is the larger deficit angle.  
Also one should use the Kerr geometry to take into account the
spin and angular momentum of the particles, but the effects on the
quantization condition of doing
so 
are 
subleading, of relative order $(s~{\rm or}~ l)\delta$.
  
\REF\csgravity{E. Witten, {\it Nucl. Phys.} {\bf B311}, 46 (1988).}

Extension of the calculation to higher orders in $G$ raises questions
on another level, since in coupling supergravity to
matter we are inevitably dealing with non-renormalizable
theories, at least perturbatively.  
The foregoing considerations, which depend only on the
behavior of the theory at low energies, should presumably hold good
independent of how the theory is ultimately cut off at very high
energies.  The non-renormalizability would seem to present a serious
barrier to calculating obstructed supersymmetry 
splittings due to self-energy, however.  We
certainly expect such corrections to arise from Feynman graphs with
non-trivial topological structure, for reasons similar to those
mentioned above in connection with nonabelian Chern-Simons theories.
Indeed, one can formulate the pure gravity theory, at least, as a
special sort of non-compact nonabelian Chern-Simons theory 
[\csgravity ] (which 
is not only renormalizable, but topological).
Perhaps for some strongly coupled supersymmetric theory the ultraviolet
behavior is sufficiently improved that the theory remains renormalizable,
non-perturbatively, even after coupling to 
supergravity.

{\bf Acknowledgments}

We wish to thank S. Trivedi for discussions which helped
stimulate this investigation.

\refout

\end